\documentclass[twocolumn,showpacs,aps,prb,amsmath,amssymb,floatfix]{revtex4}
\usepackage{graphicx}
\usepackage{dcolumn}
\usepackage{bm}

\begin{document}
\title{Resonance Kondo Tunneling through a Double Quantum Dot at Finite Bias}
\author{M.N. Kiselev$^1$, K. Kikoin$^2$ and L.W.Molenkamp$^3$}
\affiliation{
$^1$Institut f\"ur Theoretische Physik, $^3$Physikalisches Institut ($EP\;3$), Universit\"at W\"urzburg,
D-97074 W\"urzburg, Germany\\
$^2$Ben-Gurion University of the Negev, Beer-Sheva 84105, Israel}
\date{\today}


\begin{abstract}
It is shown that the resonance Kondo tunneling  through
a double quantum dot (DQD) with {\it  even} occupation and
{\it singlet} ground state may arise at a strong bias, which compensates
the energy
of singlet/triplet excitation. Using the renormalization group technique
we derive scaling equations and calculate the differential conductance as
a function of an auxiliary dc-bias for parallel DQD described by
$SO(4)$ symmetry. We analyze the decoherence effects
associated with the triplet/singlet relaxation in DQD and discuss the
shape of differential conductance line as a function of dc-bias and 
temperature.
\end{abstract}

\maketitle
\section{Introduction}
Many  fascinating  collective effects, which exist in
strongly correlated electron systems (metallic
compounds containing transition and rare-earth elements) may be
observed also in artificial nanosize devices
(quantum wells, quantum dots, etc). Moreover,  fabricated
nanoobjects provide unique possibility to create
such conditions for observation of many-particle phenomena,
which by no means may be reached in "natural" conditions.
Kondo effect (KE) is one of such phenomena.
It was found theoretically \cite{Glazr88b,Ng88}
and observed experimentally \cite{Gogo99,Cron98,Simmel99} that the charge-spin
separation in low-energy excitation
spectrum of quantum dots under strong Coulomb blockade manifests
itself as a resonance Kondo-type tunneling through a dot with
odd electron occupation ${\cal N}$ (one unpaired spin $S=1/2$).
This resonance tunneling through a quantum dot connecting two
metallic reservoirs (leads) is an analog of resonance spin scattering
in metals with magnetic impurities.
A Kondo-type tunneling arises under conditions
which do not exist in conventional metallic
compounds. 
The KE emerges as a
dynamical phenomenon in strong time dependent 
electric field \cite{Goldin}-\cite{Lopez01},
it may arise at finite frequency under light illumination 
\cite{Kex,Shab00,Fujii01}.
Even the net zero spin of isolated quantum
dot (even ${\cal N}$) is not an obstacle
for the resonance Kondo tunneling. In this case it may be
observed in double quantum dots (DQD) arranged in parallel geometry 
\cite{KA02},
in T-shaped DQD \cite{KA01,KA02,Taka}, in two-level single dots 
\cite{Izum,Hoft}
or induced by strong magnetic field
\cite{Magn,Eto00,Giul01,Pust00} whereas in conventional metals magnetic field only
suppresses the Kondo scattering. The latter
effect was also discovered experimentally \cite{Sasa,Nyg00,Wiel02}.

One of the most challenging options in Kondo physics of quantum dots 
is the possibility of controlling
the Kondo effect by creating the non-equilibrium reservoir of fermionic
excitations by means of
 strong bias $eV \gg T_K$  applied between the leads
\cite{Meir} ($T_K$ is the equilibrium Kondo temperature which
determines the energy scale of low-energy spin excitations
in a quantum dot). However, in this case the decoherence effects may prevent
the formation of a full scale Kondo resonance (see,
e.g. discussion in Refs. \onlinecite{RKW,neq,Col}). It was argued in recent
disputes that the processes, associated with the finite current through a
dot with odd ${\cal N}$ may destroy the coherence on an energy scale 
$\Gamma \gg T_K$ and thus prevent formation of a ground state Kondo singlet,
so that only the weak coupling Kondo regime is possible in strongly
non-equilibrium conditions. 

In the present paper we discuss Kondo tunneling through DQD with even 
${\cal N}$,
whose ground state is a spin singlet $|S\rangle$. It will be shown that the
Kondo tunneling through {\it excited} triplet state  $|T\rangle$  
arises at finite $eV$. In this case the ground state is stable against any
kind of spin-flip processes induced by external current, the decoherence
effects develop only in the intermediate (virtual) triplet state,  
and the estimates of decoherence rate should be revisited. 

As was noticed in Ref. \onlinecite{KA01}, 
quantum dots with even ${\cal N}$ possess
the dynamical symmetry $SO(4)$ of {\it spin rotator} 
in the Kondo tunneling regime,
provided the low-energy part of excitation spectrum 
is formed by a singlet-triplet
(ST) pair, and all other excitations are separated from the ST manifold by
a gap noticeably exceeding the tunneling rate $\gamma$. A DQD with
even ${\cal N}$ in a side-bound (T-shape) configuration
where two wells are coupled by the tunneling $v$ and only
one of them (say, $l$) is coupled to metallic
leads $(L,R)$ is a simplest system satisfying this condition
\cite{KA01}. Such system was realized experimentally in Ref.\onlinecite{mol95}.
Novel features introduced by the dynamical symmetry in Kondo tunneling
are connected with the fact that unlike the case of conventional $ SU(2)$
symmetry of spin vector ${\bf S}$, the $SO(4)$ group possesses two generators
${\bf S}$ and ${\bf P}$. The latter vector describes transitions
 between singlet
and triplet states of spin manifold (this vector is an analog of Runge-Lenz
vector describing the hidden symmetry of hydrogen atom). As was shown in   
Ref. \onlinecite{KA02}, this vector alone is responsible for Kondo tunneling through
quantum dot with even ${\cal N}$  induced
by external magnetic field.   

Another manifestation of dynamical symmetry peculiar to DQDs 
with even ${\cal N}$ is revealed in this paper.
It is shown that in the case when the ground state is singlet $|S\rangle$
and the S/T gap $\delta \gg T_K$, a Kondo resonance
channel arises under a strong bias $eV$ comparable with $\delta.$
The channel opens at
$|eV - \delta|< T_K$, and the tunneling
is determined by the
{\it non-diagonal} component
$J_{ST}=\langle T|J|S\rangle$ of effective exchange
induced by the electron tunneling through DQD (see Fig. 1 (right panel)). 

\section{Cotunneling Hamiltonian of T-shaped DQD}
The basic properties of symmetric
DQD occupied by even number of electrons ${\cal N}=2n$ under strong
Coulomb blockade in each well are
manifested already in the simplest case $n=1$, which is considered below.
Such DQD is an artificial
analog of a hydrogen molecule ${\rm H_2}$. If the inter-well Coulomb blockade
$Q$ is strong enough, one has ${\cal N}=n_l+n_r,$ $n_l=n_r=1,$
the lowest states of DQD are singlet and triplet and
the next levels are separated from ST pair by a charge transfer gap $\sim Q$.
We assume that both wells are neutral at $n_{l,r}=1$.
Then the effective inter-well exchange $I$ responsible for the singlet-triplet
splitting arises because of
tunneling $v$ between two wells, $I=v^2/Q=\delta$. It is convenient
to write the effective spin Hamiltonian of isolated DQD in the form
\begin{equation}
H_{d}=E_S|S\rangle\langle S|+\sum_{\eta}E_T|T\eta\rangle\langle T\eta| \equiv
\sum_{\Lambda=S,T\eta}E_{\Lambda}X^{\Lambda\Lambda}
\label{1.1}
\end{equation}
\noindent
where $X^{\Lambda\Lambda'}=|\Lambda\rangle\langle \Lambda'|$ is a Hubbard
configuration change operator (see, e.g., \cite{Hewson}),
$E_T=E_S+\delta$, $\eta=\pm,0$ are three projections of $S=1$ vector.
Two other terms completing the Anderson Hamiltonian, which describes
the system shown in Fig.1 (left panel), are
$$
H_{b}+H_{t}=
$$
\begin{equation}
\sum_{k\alpha\sigma}\epsilon_{k\alpha}
c^{\dagger}_{k\alpha\sigma}c_{k\alpha\sigma} +
\sum_{\Lambda \lambda }\sum_{k\alpha\sigma}\left(W^{\Lambda \lambda}_{\sigma}
c_{k\alpha\sigma}^{\dagger }X^{\lambda \Lambda }+ H.c.
\right).
\label{2.1}
\end{equation}
\noindent
The first term describes metallic electrons in the leads and the second one stands for tunneling
between the leads and the DQD. Here $\alpha=L,R$ marks
electrons in the left and right lead, respectively,
the bias $eV$ is applied to the left lead, so that the chemical potentials are
$\mu_{FL}=\mu_{FR}+eV$,~ $W^{\Lambda \lambda}_{\sigma}$ 
is the tunneling amplitude for the well $l$ (left),
$|\lambda\rangle$ are one-electron states of DQD,
 which arises after escape of an electron with spin projection
$\sigma$ from DQD in a state $|\Lambda\rangle$.
\begin{figure}
\includegraphics[width=0.2\textwidth]{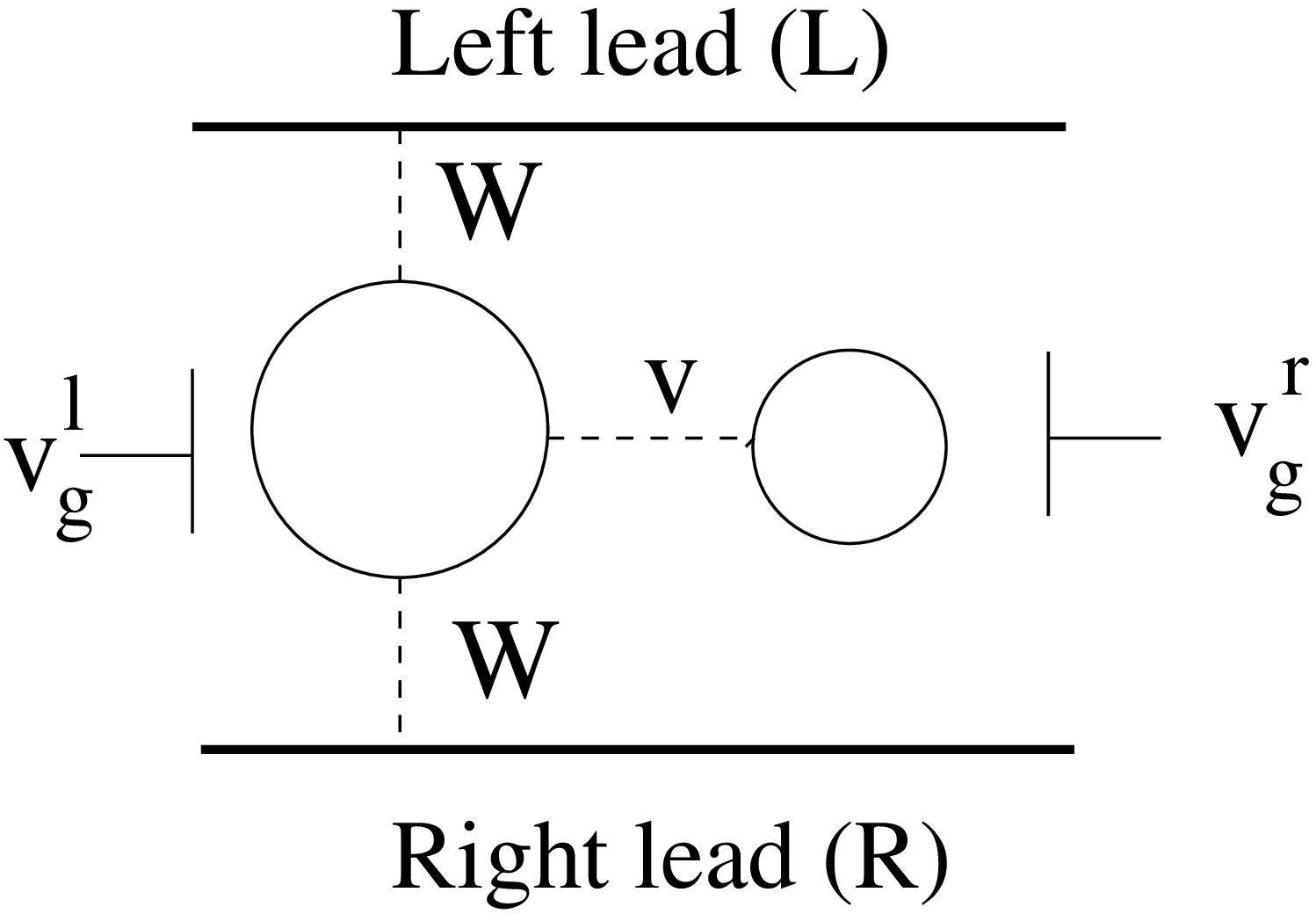}
\includegraphics[width=0.2\textwidth]{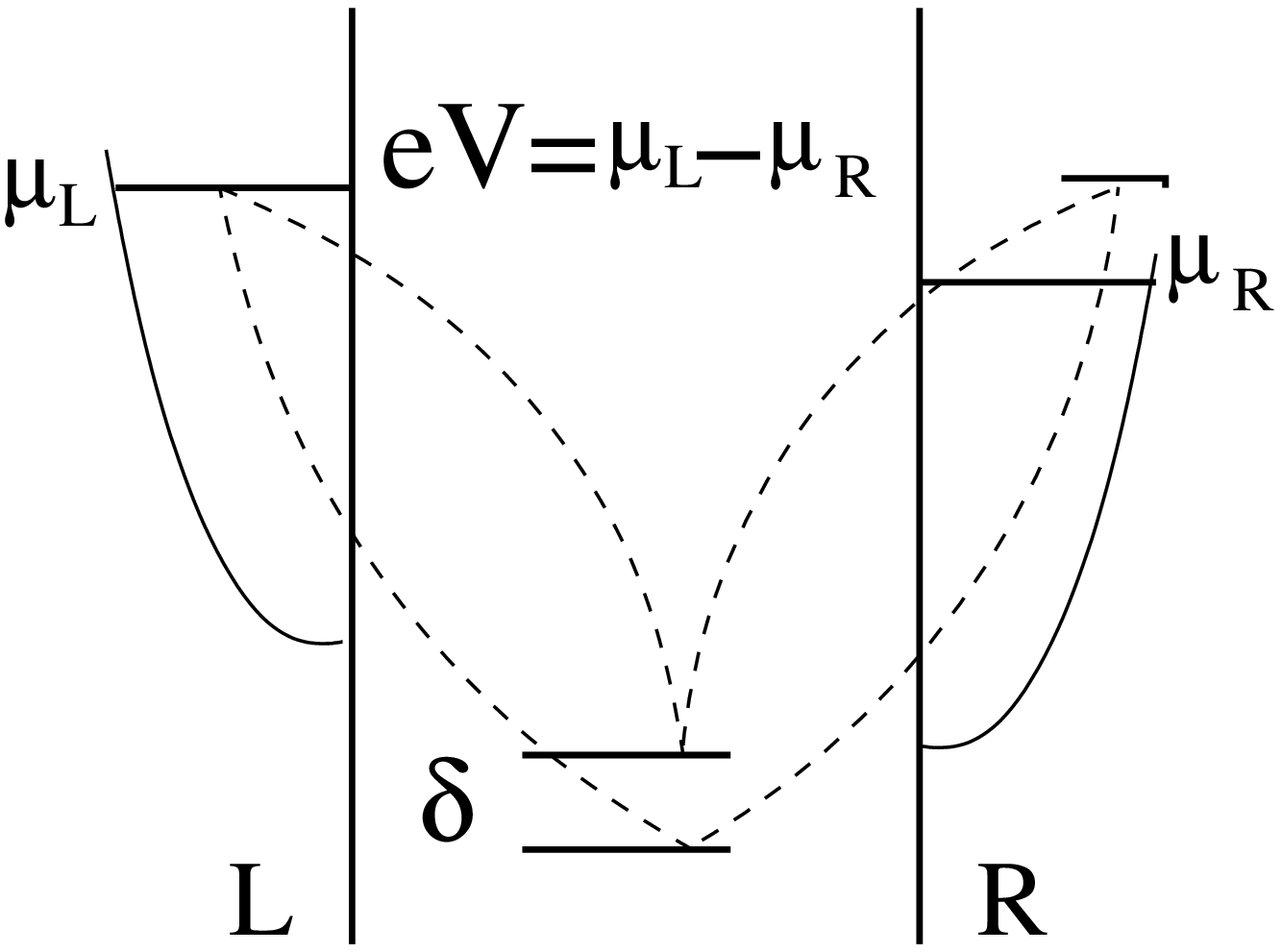}
\caption{Left panel: Double quantum dot in a side-bound configuration.
Right panel: cotunneling processes
in biased DQD responsible for the resonance Kondo tunneling.}
\end{figure}
We solve the problem in a Schrieffer-Wolff (SW) limit
\cite{Hewson}, when the activation energies
$|E_\Lambda-E_\lambda-\mu_{F\alpha}|$ and Coulomb blockade energy
$Q$ are essentially larger then the tunneling rate $\gamma$, and
charge fluctuations are completely suppressed both in the ground
and excited state of DQD.  In this limit one may start with the SW
transformation, which projects out charge excitations. We confine
ourselves with the bias $eV \lesssim \delta \ll D$, where $D$ is the
width of the electrons in the leads, so the leads are considered
in the SW transformation as two independent quasi equilibrium
reservoirs (cf. \cite{Goldin,Kam00}). As is shown in Ref.
\cite{KA01}, the SW transformation being applied to a spin rotator
results in the following effective spin Hamiltonian

\begin{equation}
H_{int}=\sum_{\alpha\alpha^\prime}[
(J^{TT}_{\alpha\alpha^\prime}{\bf S} +  J^{ST}_{\alpha\alpha^\prime}
{\bf P})\cdot {\bf s}_{\alpha\alpha^\prime}
+J^{SS}_{\alpha\alpha^\prime}X^{SS}n_{\alpha\alpha^\prime}]
\label{2.2}
\end{equation}
\noindent
Here ${\bf s}_{\alpha\alpha^\prime}$$=$$
\sum_{kk'}c^{\dagger}_{k\alpha\sigma}\hat{\tau}c_{k'\alpha'\sigma'}$,~
$n_{\alpha\alpha^\prime}$$=$$
\sum_{kk'}c^{\dagger}_{k\alpha\sigma}\hat{1}c_{k'\alpha'\sigma}$,
$\hat{\tau}$,~ $\hat{1}$ are the Pauli matrices and unity matrix respectively.
The effective exchange constants are
$$
J^{\Lambda\Lambda'}_{\alpha\alpha^\prime} \approx
\frac{W^{\Lambda\lambda}_\sigma W^{*\lambda\Lambda'}_\sigma}{2}
\left( \frac{1}{\epsilon_{F\alpha}-E_S/2} +
\frac{1}{\epsilon_{F\alpha^\prime}-E_S/2} \right).
$$
In this approximation the small differences between singlet and
triplet states are neglected. Besides,
$J_{\alpha\alpha'}^{\Lambda\Lambda'}\sim I$ in real DQD.

Two vectors ${\bf S}$ and ${\bf P}$  with spherical components
\begin{eqnarray}
\nonumber
S^+  = \sqrt{2}\left(X^{10}+X^{0-1}\right),~
S^-  =  \sqrt{2}\left(X^{01}+X^{-10}\right),\\
\nonumber
S_z  =  X^{11}-X^{-1-1},\,\;\;\;
P_z  =  -\left(X^{0S}+X^{S0}\right),\\
P^+  = \sqrt{2}\left(X^{1S}-X^{S-1}\right),
P^-  = \sqrt{2}\left(X^{S1}-X^{-1S}\right).
\label{SP}
\end{eqnarray}
\noindent
obey the commutation relations of $o_4$ algebra
\[
[S_j,S_k]  = ie_{jkl}S_l,~[P_j,P_k]=ie_{jkl}S_l,~ [P_j,S_k]=ie_{jkl}P_l
\]
\noindent
($j,k,l$ are Cartesian coordinates, $e_{jkl}$ is a Levi-Civita tensor).
These vectors are orthogonal, ${\bf S\cdot P} = 0,$ and the Casimir operator
is ${\bf S}^2+ {\bf P}^2 =3.$ Thus, the singlet state is involved
in spin scattering via the components of the vector ${\bf P}$.

We use $SU(2)$-like semi-fermionic representation for $S$
operators \cite{popov,kis00}
$$
S^+  =   \sqrt{2}(f_0^\dagger f_{-1}+f^\dagger_{1}f_0),\;\;\;
S^-  =  \sqrt{2}(f^\dagger_{-1}f_0+ f_0^\dagger f_{1}),
$$
\begin{equation}
S_z  =  f^\dagger_{1}f_{1}-f^\dagger_{-1}f_{-1},
\label{spin}
\end{equation}
\noindent where $f^\dagger_{\pm 1}$ are creation operators for
fermions with spin ``up'' and ``down'' respectively, whereas $f_0$
stands for spinless fermion \cite{popov,kis00}. This
representation can be generalized for $SO(4)$ group by introducing
another spinless fermion $f_s$ to take into consideration the
singlet state. As a result, the $P$-operators are given by the
following equations:
$$
P^+ =  \sqrt{2}(f^\dagger_{1} f_s -  f_s^\dagger f_{-1}),\;\;\;
P^- =  \sqrt{2}(f_s^\dagger f_{1} - f^\dagger_{-1}f_s),
$$
\begin{equation}
P^z  =  -( f_0^\dagger f_s + f_s^\dagger f_0).
\label{proj}
\end{equation}
%
\noindent
The Casimir operator ${\bf S}^2+{\bf P}^2=3$ transforms to the local constraint
$$\sum_{\Lambda=\pm,0,s}f^\dagger_\Lambda f_\Lambda=1.$$
The final form of the spin cotunneling Hamiltonian is
\begin{eqnarray}
&&H_{int}=\sum_{kk',\alpha\alpha'=L,R}J^S_{\alpha\alpha'}
f_s^\dagger f_s c^\dagger_{k\alpha\sigma}c_{k'\alpha'\sigma}
\label{hint} \\
&&+\sum_{kk',\alpha\alpha'\Lambda\Lambda'}\left(J^T_{\alpha\alpha'}
\hat S^d_{\Lambda \Lambda'}
+J^{ST}_{\alpha\alpha'}\hat P^d_{\Lambda \Lambda'}\right)\tau^d_{\sigma\sigma'}
c^\dagger_{k\alpha\sigma}c_{k'\alpha'\sigma'}f_\Lambda^\dagger f_{\Lambda'}
\nonumber
\end{eqnarray}
\noindent
where $\hat S^d$ and $\hat P^d$ ($d$$=$$x$,$y$,$z$) are $4\times 4$
matrices  defined by relations (\ref{SP}) - (\ref{proj})
and $J^S=J^{SS}$, $J^T=J^{TT}$ and $J^{ST}$ are singlet, triplet and   
singlet-triplet
coupling SW constants, respectively. 

The cotunneling in the ground singlet state is described by the first term
of the Hamiltonian (\ref{hint}),
and no spin flip processes accompanying the electron
transfer between the leads emerge in this state. However, the last term in 
(\ref{hint}) links the singlet ground state with the excited triplet and opens
a Kondo channel. In equilibrium this channel is ineffective, because 
the incident electron should have the energy $\delta$ to be able to
initiate spin-flip processes. We will show in the next section that the
situation changes radically, when strong enough external bias is applied.

\section{Kondo singularity in tunneling through DQD at finite bias}

We deal with the case, which was not met in the previous
studies of non-equilibrium Kondo tunneling. The ground state of
the system is singlet, and the Kondo tunneling in equilibrium is
quenched at $T \sim \delta$. Thus, the elastic Kondo tunneling
arises only provided $T_K \gg\delta$ in accordance with the theory
of two-impurity Kondo effect \cite{KA01,Varma}. However, the
energy necessary for spin flip may be donated by external electric
field $eV$ applied to the left lead, and in the opposite limit
$T_K \ll\delta$ the elastic channel emerges at $eV\approx \delta$.
The processes responsible for resonance Kondo cotunneling at finite
bias are shown in Fig. 1 (left panel).

In conventional spin $S=1/2$ quantum dots the Kondo regime out of
equilibrium is affected by spin relaxation and decoherence
processes, which emerge at $eV \gg T_K$ (see, e.g.,
\cite{Kam00,RKW,neq,Col}). These processes appear in the same
order as Kondo co-tunneling itself, and one
should use the non-equilibrium perturbation theory (e.g., Keldysh
technique) to take them into account in a proper way. In our case
these effects are expected to be weaker, because the nonzero spin state is
involved in Kondo tunneling only as an intermediate virtual state
arising due to S/T transitions induced by the second term in
the Hamiltonian (\ref{2.2}), which contains vector ${\bf P}$. The
non-equilibrium repopulation effects in DQD are weak as well 
(see next section, where the nonequilibrium effects are discussed
in more details).

Having this in mind, we describe Kondo tunneling through DQD at
finite $eV\lesssim \delta$ within the quasi-equilibrium perturbation
theory in a weak coupling regime (cf. the quasi-equilibrium
approach to description of decoherence rate at large $eV$ in Ref.
\onlinecite{RKW}). To develop the perturbative approach for $T>T_K$ we
introduce the temperature Green's functions (GF) for electrons in
a dot, ${\cal G}_{\Lambda}(\tau)=-\langle T_\tau f_\Lambda(\tau)
f^\dagger_\Lambda(0)\rangle,$ and GF for the electrons in the left
(L) and right (R) lead, $G_{L,R}(k,\tau)=-\langle T_\tau c_{L,R
\sigma}(k,\tau) c^\dagger_{L,R\sigma}(k,0)\rangle$. Performing a
Fourier transformation in imaginary time for bare GF's, we come to
following expressions:
$$
G^0_{k\alpha}(\epsilon_n)=(i\epsilon_n -\epsilon_{k\alpha}+\mu_{L,R})^{-1},
$$
$$
{\cal G}^0_{\eta}(\omega_m)=(i\omega_m -E_T)^{-1},\;\;\; \eta=-1,0,1
$$
\begin{equation}
{\cal G}^0_{s}(\epsilon_n)=(i\epsilon_n -E_S)^{-1},
\label{GF}
\end{equation}
\noindent with $\epsilon_n=2\pi T(n+1/2)$ and $\omega_m=2\pi
T(m+1/3)$ \cite{popov,kis00}. The first leading and next to
leading parquet diagrams are shown on Fig.2.
\begin{figure}
\includegraphics[width=0.4\textwidth]{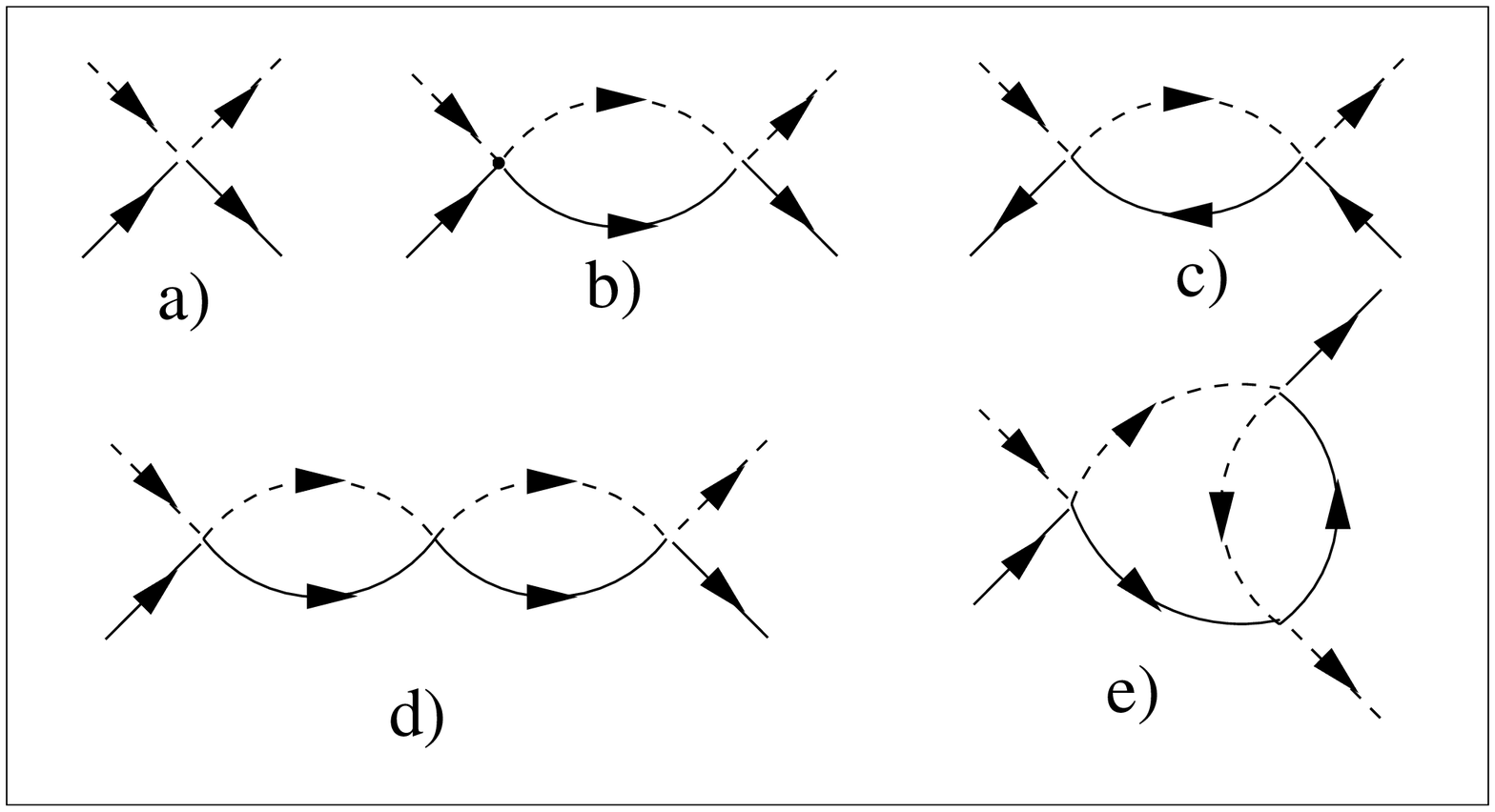}
\caption{Leading (b,d) and next to leading (c,e) parquet diagrams
determining renormalization of $J^S (a)$.
Solid lines denote electrons in the leads.
Dashed lines stand for electrons in the dot.}
\end{figure}
Corrections to the singlet vertex $\Gamma(\omega,0; \omega',0)$
are calculated using an analytical continuation of GF's to the
real axis $\omega$ and taking into account the shift of the
chemical potential in the left lead. 
Since the electron from the left lead tunnels 
into the empty state in the right lead
separated by the energy $eV$, we have to put $\omega=eV$, $\omega'=0$ in the 
final expression for $\Gamma(\omega,0; \omega',0)$. 
Thus, unlike conventional Kondo effect
 we deal with the vertex at finite frequency
$\omega$ similarly to the problem considered in Ref. \onlinecite{RKW}. 
We assume that the leads remain in 
equilibrium under applied bias and neglect the relaxation processes 
in the leads (``hot'' leads).
In a weak coupling regime
$T>T_K$ the leading non-Born contributions to the tunnel current
are determined by the diagrams of Fig. 2 b-e.

The effective vertex shown in Fig. 2b
is given by the following equation
\begin{equation}\label{sing}
\Gamma_{LR}^{(2b)}(\omega)=J_{LL}^{ST}J_{LR}^{TS}\sum_{\bf k}
\frac{1-f(\epsilon_{kL}-eV)}
{\omega - \epsilon_{kL} + \mu_L - \delta}
\end{equation}
\noindent Changing the variable $\epsilon_{kL}$ for
$\epsilon_{kL}-eV$ one finds that 
$$\Gamma_{LR}^{(2b)}(\omega=eV)\sim
J_{LL}^{ST}J_{LR}^{TS} \nu \ln(D/{\rm max}\{
(eV-\delta),T\}).$$ 
Here $D \sim \varepsilon_F$ is a cutoff energy
determining effective bandwidth, $\nu$ is a density of states on a
Fermi level and $f(\varepsilon)$ is the Fermi function. Therefore,
under  condition $|eV-\delta|\ll \max[eV,\delta]$ this correction
does not depend on $eV$ and becomes quasielastic.

Unlike the diagram Fig. 2b,
its "parquet counterpart" term Fig. 2c contains  $eV+\delta$
in the argument of the Kondo logarithm:
\begin{equation}
\Gamma_{LR}^{(2c)}(\omega)=J_{LL}^{ST}J_{LR}^{TS}\sum_{\bf k}
\frac{f(\epsilon_{kL}-eV)}
{\omega - \epsilon_{kL} + \mu_L + \delta}
\end{equation}
\noindent At $eV\sim\delta \gg T$ this contribution is
estimated as 
$$\Gamma_{LR}^{(2c)}(eV)\sim
J_{LL}^{ST}J_{LR}^{TS}\nu \ln~(D/(eV+\delta)) \ll
\Gamma_{LR}^{(2b)}(eV).$$

Similar estimates for diagrams Fig.2d and 2e give
$$\Gamma_{LR}^{(2d)}(\omega)\sim
J_{LL}^{ST}J_{LL}^{T}J_{LR}^{TS}\nu^2\ln^2~(D/{\rm max}\{\omega, (eV-\delta),T\})
$$
$$\Gamma_{LR}^{(2e)}(\omega)\sim
J_{LL}^{ST}J_{LL}^{T}J_{LR}^{TS}\nu^2\ln~(D/{\rm max}\{\omega, (eV-\delta),T\})\times
$$
\begin{equation}
\times \ln~(D/{\rm max}\{\omega, eV, T\}.
\label{2nd}
\end{equation}
\noindent
Then $\Gamma_{LR}^{(2e)}(\omega)\ll\Gamma_{LR}^{(2d)}(\omega)$
at $eV \to \delta$.

Thus, the
Kondo singularity is restored in non-equilibrium conditions
where the electrons in the left lead acquire additional energy in external
electric field, which compensates the
energy loss $\delta$ in a singlet-triplet excitation. The leading sequence of
most divergent diagrams degenerates in this case from a parquet to a
ladder series.

Following the poor man's scaling approach, we derive the system of coupled
renormalization group (RG) equations for (\ref{hint}). The equations for LL
co-tunneling are:
\begin{equation}
\frac{d J^T_{LL}}{d \ln D}=-\nu (J_{LL}^T)^2,\;\;\;
\frac{d J^{ST}_{LL}}{d \ln D}=-\nu J_{LL}^{ST} J_{LL}^T,
\label{ll}
\end{equation}
\noindent
The scaling equations for $J_{LR}^\Lambda$ are as follows:
$$
\frac{d J^T_{LR}}{d \ln D}=-\nu J_{LL}^T J_{LR}^T,\;\;\;
\frac{d J^{ST}_{LR}}{d \ln D}=-\nu J_{LL}^{ST} J_{LR}^T,
$$
\begin{equation}
\frac{d J^S_{LR}}{d \ln D}=\frac{1}{2}\nu\left(J_{LL,+}^{ST}J_{LR,-}^{TS}
+\frac{1}{2}J_{LL,z}^{ST}J_{LR,z}^{TS}\right).
\label{rg}
\end{equation}
\noindent
One-loop diagrams corresponding to the poor man's scaling procedure are shown
in Fig. 3.
To derive these equations we collected only terms
$\sim (J^T)^n\ln^{n+1}(D/T)$ neglecting
contributions containing $\ln[D/(eV)]$. The analysis of RG equations beyond
the one loop approximation
will be published elsewhere.
\begin{figure}
\includegraphics[width=0.4\textwidth]{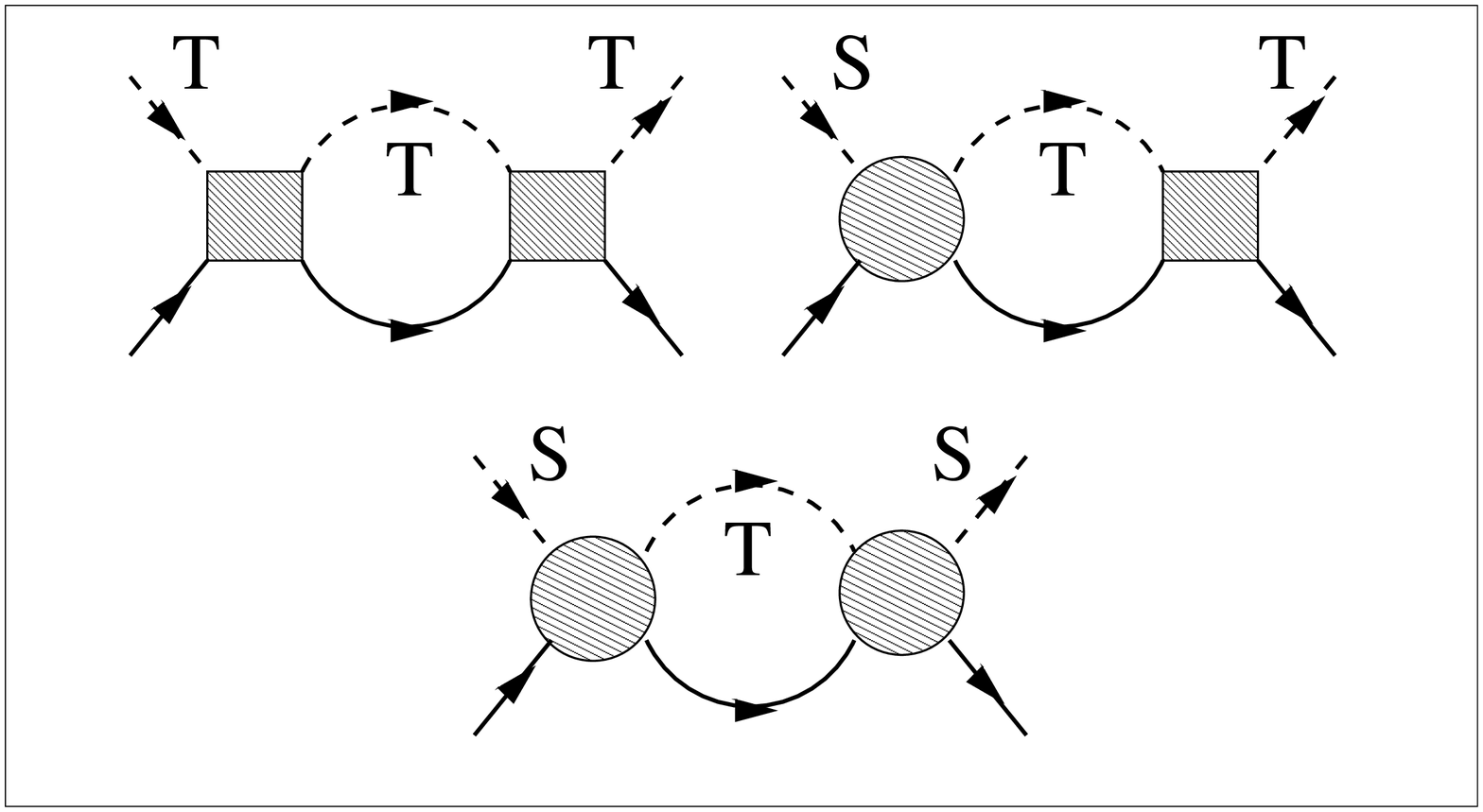}
\caption{Irreducible diagrams contributing to RG equations. Hatched boxes and circles stand for
triplet-triplet and singlet-triplet vertices respectively. Notations for lines are the same as in Fig.2}
\end{figure}
The solution of the system (\ref{rg}) reads as follows:
$$
J^T_{\alpha,\alpha'}=\frac{J^T_{0}}{1-\nu J^T_{0}\ln(D/T)},\;\;\;
J^{ST}_{\alpha,\alpha'}=\frac{J^{ST}_{0}}{1-\nu J^T_{0}\ln(D/T)},
$$
\begin{equation}
J^S_{LR}=J^S_{0}-\frac{3}{4}\nu (J^{ST}_0)^2\frac{\ln(D/T)}{1-\nu J^T_{0}\ln(D/T)}.
\label{srg}
\end{equation}
\noindent
Here $\alpha=L$, $\alpha'=L,R$.
One should note that the Kondo temperature is determined by 
triplet-triplet processes only in spite of the fact that the ground state
is singlet.
One finds from (\ref{srg}) that $T_K=D\exp[-1/(\nu J^T_0)]$. 
This temperature is 
noticeably smaller than the "equilibrium" Kondo temperature $T_{K0}$, 
which emerges in tunneling
through  triplet channel in the ground state, namely   
$T_K\approx T_{K0}^2/D$. The reason for this difference is
the reduction of usual parquet equations for $T_K$ 
to a simple ladder series. In this respect our case differs also from
conventional Kondo effect at strong bias \cite{RKW}, where the 
non-equilibrium Kondo temperature $T^*\approx T_{K0}^2/eV$ arises. 
In our model the finite
bias does not enter $T_K$ because of the compensation 
$eV\approx \delta$ in spite of the fact that we take the argument 
$\omega=eV$ in the vertex (\ref{sing}). 
\begin{figure}
\includegraphics[width=0.4\textwidth]{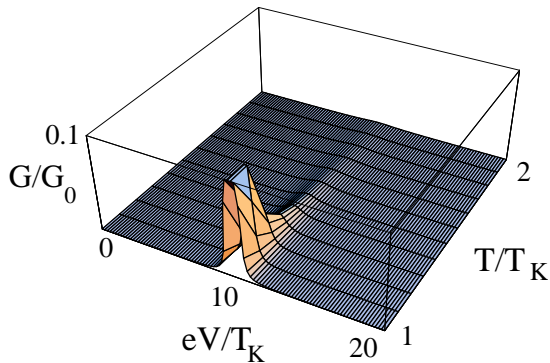}
\caption{The Kondo conductance as a function of dc-bias $eV/T_K$ and $T/T_K$.
The singlet-triplet splitting $\delta/T_K=10$.}
\end{figure}
The differential conductance $G(eV,T)/G_0\sim |J^{ST}_{LR}|^2$ 
(cf. Ref. \onlinecite{KNG}) is the universal function of
two parameters $T/T_K$ and $eV/T_K$,~ $G_0=e^2/\pi \hbar$:
\begin{equation}
G/G_0 \sim \ln^{-2}\left(\max[(eV-\delta),T]/T_K\right)
\label{dcond}
\end{equation}
Its behavior as a function of bias and temperature is shown in Fig. 4. 
It is seen from this picture that the resonance tunneling "flashes" at
$eV \sim \delta$ and dies away out of this resonance. In this picture the 
decoherence effects are not taken into account, and it stability against
various non-equilibrium corrections should be checked.

\section{Decoherence effects}
 
We analyze now the decoherence rate $\hbar/\tau_d$
associated with T/S transition relaxation induced by 
cotunneling. The calculations are performed  
in the same order of the
perturbation theory as it has been done for the vertex
renormalization (see Figs. 2 and 3). The details of the 
calculation scheme are presented in Appendix. 

To estimate the decoherence effects, one should calculate the 
decay of the triplet state  or in other terms to find the imaginary part   
of the retarded 
self energy of triplet semi-fermion propagators at actual frequency
[see discussion before Eq. (\ref{sing})], 
$\hbar/\tau_d = -2 {\rm Im} \Sigma^R_T (\omega)$.  
The 2nd and 3rd order diagrams
determining $\hbar/\tau_d$ are shown in Fig.5(a-d).
Two leading terms given by the diagrams of  Fig.5(a,b) 
describe the damping of triplet excitation due to its inelastic relaxation 
to the ground singlet state. These terms 
are calculated in Appendix (see Eqs. (\ref{eq10}), (\ref{eq10a})). 
One finds from these equations that the relaxation rate associated 
with ST transition is
\begin{equation}\label{relst}
1/\tau_d^{ST} \sim\left(J^{ST}/D \right)^2 {\rm
max}[eV,\omega,T_K]. 
\end{equation}
It should be noted that for
corrections associated with LL (RR) diagrams (Fig.5a), describing
co-tunneling processes on a left (right) lead, the use of
quasi-equilibrium technique  is fully justified when the leads
themselves are  in thermal equilibrium. 
We are interested in the zero frequency damping at resonance 
$eV\approx \delta$. Neglecting the small difference between $J^T$ and
$J^{ST}$ (see \cite{KA01}), we also take $J^T \approx J^{ST} =J$. Thus the
$T\to S$ spin relaxation effect (\ref{relst}) 
does not contain logarithmic enhancement
factor in the lowest order. It is estimated as
\begin {equation}\label{relsto}
1/\tau_d^{ST}\sim (eV)\cdot (J/D)^2 \approx J^3/D^2~. 
\end{equation}

The repolulation of triplet state as a function of external 
bias is controlled by the occupation number for triplet state 
modified by the bias $eV$. The latter, in turn, depends on 
the modified exchange splitting $\delta^*$ given by solution 
of the equation
\begin{equation}
\delta^* - \delta= Re \Sigma^R(\delta^*, eV, T).
\end{equation}
The $Re \Sigma^R$ (Fig.5(a,b)) is given by
\begin{equation}
Re \Sigma^{R(2)}_{TST}(\omega, eV,T)= -a_2 \left(\frac{J}{D}\right)^2 \omega 
\ln\left(\frac{D}{{\rm max}[\omega, eV, T]}\right)
\end{equation} 
where $a_2 \sim 1$ is a numerical coefficient. As it is seen, the perturbative
equation for $Re \Sigma^R$ is beyond the scope of leading-log approximation.
As a result,  
$\delta^*(eV) -\delta\ll \delta$ and repopulation of the triplet state 
is exponentially small. The corresponding factor in the occupation number is
\begin{equation}
P_t(eV)=\exp\left(-\delta^*(eV)/T\right)
\end{equation}

The effects of repopulation become important only at 
$eV \gg \delta$ when $|\delta^*-\delta|\sim \delta$. In that case
the quasi equilibrium approach is not applicable and one should 
start with the Keldysh formalism \cite{RKW,neq}. This regime is definitely not realized in conditions considered above.

Next 2nd order contribution is the damping of triplet state itself
given by Eqs. (\ref{eq22}), (\ref{eq22a}). It is seen from these equations
that this damping is of threshold character:
\begin{equation}
1/\tau_d^{TT} \sim ( J/D)^2 (\omega -\delta)\theta (\omega -\delta)~,
\end{equation}
where $\theta (\omega)$ is a Heaviside step function. 
These processes emerge only at
$\omega>\delta$, so unlike the conventional
case \cite{RKW} they are not dangerous.

Corresponding contribution to $Re \Sigma^R$ casts the form
\begin{equation}
Re \Sigma^{R(2)}_{TTT}= -b_2 \left(\frac{J}{D}\right)^2 
(\delta- \omega) 
\ln\left|\frac{D}{{\rm max}[(\delta-\omega), T]}\right|,
\end{equation}
where $b_2\sim 1$.

\begin{figure}
\includegraphics[width=0.4\textwidth]{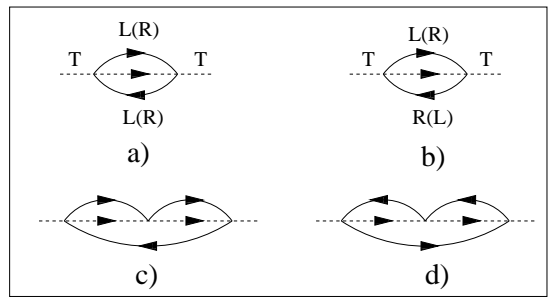}
\caption{Leading diagrams (a-d) for $\hbar/\tau_{d}$ (see text). Dashed line in the self-energy part
stands for the singlet state of a two-electron configuration in the dot.}
\end{figure}
Next, one have to check whether the higher order logarithmic corrections   
modify the estimate (\ref{relsto}). These corrections 
start with the 3-rd order terms
shown in Fig. 5(c,d). Straightforward calculations lead to  $eV
(J/D)^3\ln(D/eV)$ correction (see the first term Eq. \ref{eq17}).
This leading term like the 2-nd order term originates from $T\to S$ 
spin relaxation 
processes. All other contributions are either of threshold character, or
vanish at $\omega \to 0$. 
As a result, the estimate
$$
\hbar/\tau_{d}\sim eV (J^{ST}_0/D)^2[1+O(\nu J\ln(D/(eV))]
$$
holds.  The topological structure (sequence of intermediate singlet
and triplet states and cotunneling processes in the left and right lead) 
in perturbative corrections for the triplet self energy part is
different from those for the
singlet-singlet vertex (see Appendix). Namely, the leading (ladder)
diagrams for the vertex contain
maximal possible number of intermediate triplet states, whereas
the higher order non-threshold log-diagrams
for the self energy part must contain at least one intermediate
singlet state.
As it is seen from the Appendix (Eq.(\ref{eq14})-(\ref{eq17})), 
the higher order
contributions to $Im  \Sigma_T(\omega)$ are not universal
and the coefficients in front of log have sophisticated
frequency dependence. As a result, the perturbative series
for triplet self-energy part can not be collected in parquet structures
and remain beyond the leading-log approximation discussed in the Section
III.
There is no strong enhancement of the 2nd order term in $SO(4)$
spin rotator model in contrast to $SU(2)$ case discussed in \cite{RKW}.
As was pointed out above, the main
reason for  differences in estimates of coherence rate is that in case
of
QD with odd ${\cal N}$, the Kondo singlet develops in the ground state
of
the dot, and decoherence frustrate this ground state. In DQD with even
${\cal N}$ the triplet spin state arises only as a virtual state in 
cotunneling processes, and our calculations demonstrate explicitely
that   
decoherence effects in this case are essentially weaker.

The 3-rd order correction to $Re \Sigma$ is given by 
\begin{equation}
Re \Sigma^{R(3)}(\omega)\sim \left(\frac{J}{D}\right)^3
\omega \ln^2\left(\frac{D}{\omega}\right)
\end{equation}
(see Appendix). This correction also remains beyond the leading-log approximation.

Thus we conclude that the decoherence effects are not destructive for
Kondo tunneling through T-shaped DQD, i.e. the $T_K\gg \hbar/\tau_d$
is valid provided 
\begin{equation}\label{ineq}
\delta(\delta/D)^2\ll T_K\ll \delta. 
\end{equation}
This interval is wide enough because $\delta/D \ll 1$ in the Anderson model.

The same calculation procedure may be repeated in Keldysh technique. 
It is seen
immediately that in the leading-log approximation 
the off-diagonal terms in Keldysh matrix are not changed
in comparison with equilibrium distribution functions because of the 
same threshold character of repopulation processes, so 
in the leading approximation the key diagram 
Fig. 5b (determining L-R current through the dot) calculated in
Keldysh technique remains the same as (\ref{eq10})-(\ref{eq10a}). 

In fact,  repopulation effects 
result in asymmetry of the Kondo-peak similar to that
in Ref. \onlinecite{neq} due to the threshold character of ${\rm Im}
\Sigma_{TTT}$ (see Appendix). This asymmetry becomes noticeable
at $eV\gg \delta$, where our quasi equilibrium approach fails, 
but this region is beyond our interest, because the bias-induced
Kondo tunneling is negligible at large biases (see Fig. 4).

\section{Concluding remarks}

We have shown in this paper that the tunneling through DQD with even 
${\cal N}$ with singlet ground state and triplet excitation divided 
by the energy gap $\delta\gg T_K$ from the singlet state 
exhibits a peak in
differential conductance at $eV\approx \delta$ (Fig. 4). 
This result is in striking
contrast with the zero bias anomaly (ZBA) at  $eV\approx 0$  which arises in
the opposite limit, $\delta < T_K$. In the latter case the Kondo screening
is quenched at energies less than $\delta$, so the ZBA has a form of a dip
in the Kondo peak (see \cite{Hoft} for detailed explanation of this effect).  

In this case strong external bias initiates the Kondo effect in DQD,
whereas in a conventional situation (QD with odd ${\cal N}$ spin 1/2 
in the ground state) strong enough bias is destructive for Kondo tunneling.
We have shown that the principal features of Kondo effect in this specific
situation may be captured within a  quasi equilibrium approach. 
The scaling equations (\ref{rg}), (\ref{srg}) can also be derived in
Schwinger-Keldysh formalism (see Refs. \onlinecite{kis00,neq}) by
applying the ``poor man's scaling'' approach directly to the dot conductance
\cite{Goldin}.

Of course, our RG approach is valid only in the weak coupling regime. 
Although in our case the limitations imposed by decoherence effects are more liberal
than those existing in conventional QD, they apparently prevent the full
formation of the Kondo resonance. To clarify this point one has to use
a genuine non-equilibrium approach, and we hope to do it in forthcoming
publications.

One should mention yet another possible experimental realization
of resonance Kondo tunneling driven
by external electric field. 
Applying the alternate field $V=V_{ac}\cos(\omega t)$ to the
parallel DQD, one takes into consideration two effects, 
namely (i) enhancement of
Kondo conductance by tuning the amplitude of ac-voltage 
to satisfy the condition
$|eV_{ac}-\delta| \ll T_K$ and (ii) spin decoherence effects due to 
finite decoherence rate  \cite{Goldin}.
One can expect that if the decoherence rate $\hbar/\tau \gg T_K,$
\begin{equation}
G_{peak}/G_0 \sim \ln^{-2}\left(\hbar/\tau T_K\right)
\end{equation}
\noindent
whereas in the opposite limit $\hbar/\tau \ll T_K$,

\begin{equation}
G_{peak}=\overline{G(V_{ac}\cos[\omega t])}
\end{equation}
\noindent
is averaged over a period of variation of ac bias. In this case the estimate
(\ref{dcond}) is also valid.

In conclusion, we have provided the first example of Kondo effect, which
exists {\it only} in non-equilibrium conditions. It is driven by external
electric field in tunneling through a
quantum dot with even number of electrons, when the low-lying states are
those of spin rotator. This is not too exotic situation because as a rule,
a singlet ground state implies a triplet excitation. If the
ST pair is separated by a gap from other excitons, then
tuning the dc-bias in such a way that applied voltage compensates
the energy of triplet excitation,
one reaches the regime of Kondo peak in conductance.
This theoretically predicted effect
can be observed in dc- and ac-biased double quantum dots in parallel geometry.
\bigskip\\
\section*{ACKNOWLEDGMENTS}
This work is partially supported (MK) by the European Commission under LF project: Access
to the Weizmann
Institute Submicron Center (contract number: HPRI-CT-1999-00069). The authors
are grateful to Y. Avishai,
A. Finkel'stein, A.Rosch and M.Heiblum
for numerous discussions. The financial support of the Deutsche
Forschungsgemeinschaft (SFB-410) is
acknowledged. The work of KK is supported by ISF grant.
\begin{center}
\appendix{\bf APPENDIX }
\medskip\\
\end{center}
\setcounter{equation}{0}
\renewcommand{\theequation}{A.\arabic{equation}}
We calculate perturbative corrections for 
$\Sigma(\omega)$ by performing analytical continuation of $\Sigma(i\omega_n)$ into upper half-plane of
$\omega$. The parameter of perturbation theory is $\nu J\ll 1$ where
$\nu$ denotes the density 
of states for conduction electrons at the Fermi surface. 

The 2-nd order self energies have following structure (the indices $T$ and
$ST$ in exchange vertices are temporarily omitted): 
\begin{widetext}
\begin{equation}
\Sigma^{(2)}(i\omega_n)\sim J^2 T^2
\sum_{\omega_1\omega_2}\sum_{\bf k_1, k_2}G^0(-i\omega_1,-{\bf k}_1)
G^0(i\omega_2,{\bf k}_2){\cal G}^0(i\omega_n+i\omega_1+i\omega_2)
\label{eq5}
\end{equation}
\end{widetext}
The Green functions (GF) are defined in Eq. (\ref{GF}). 
Performing summation over 
Matsubara frequencies $\omega_1,\omega_2$  and replacing the summation over ${\bf k_1,k_2}$ 
by integration
over $\xi_1,\xi_2$ in accordance with standard procedure, we come to following expression
\begin{widetext}
\begin{equation}
\Sigma^{(2)}(i\omega_n)\sim \frac{1}{2}\left(J \nu\right)^2
\int_{-D}^{D} d\xi_1\int_{-D}^{D} d\xi_2
\frac{\displaystyle\left[\tanh\left(\frac{\lambda}{2T}\right)-\tanh\left(\frac{\xi_2}{2T}\right)\right]
\left[\tanh\left(\frac{\xi_1}{2T}\right)-\tanh\left(\frac{\xi_2-\lambda}{2T}\right)\right]}
{i\omega_n +\xi_2 -\xi_1 -\lambda_{S,T}}
\label{eq6}
\end{equation}
\end{widetext}
Here we assumed that conduction electron's band has a width $W=2D$, $\epsilon_F\sim D$ 
and $\nu =1/D$ 
in order to simplify our calculations. This assumption is sufficient for log-accuracy of our theory.
The Lagrange multipliers $\lambda_{S,T}$ are different for singlet (triplet) GF, namely 
$\lambda_S=E_S$ and $\lambda_T=E_T+i\pi T/3$

To account for decoherence effects in the same order of perturbation theory 
as we have done  for the vertex corrections, we focus on the 
self-energy (SE) part of triplet GF. This SE has to be plugged in back to 
a semi-fermionic propagator 
to provide a self-consistent treatment of the problem.
We denote the self-energy parts associated with singlet/triplet and
triplet/triplet transitions as $\Sigma°°°_{TST}$ and $\Sigma_{TTT}$ respectively. 

To prevent double occupancy of singlet/triplet states we take the limit    
$Re[\lambda_{S,T}] \gg T$ in the numerator of Eq. (\ref{eq6}). 
As a result, Eq.(\ref{eq6}) casts the form
\begin{widetext}
\begin{equation}
\Sigma^{(2)}(i\omega_n)\sim\left(J \nu\right)^2
\int_{-D}^{D} d\xi_1\int_{-D}^{D} d\xi_2
\frac{n(\xi_2)(1-n(\xi_1))}
{i\omega_n +\xi_2 -\xi_1 -\lambda_{S,T}}
\label{eq7}
\end{equation}
\end{widetext}
Since all spurious states are ``frozen out'' we can put $\tilde \lambda_S=0$ 
and $\tilde\lambda_T=\delta=E_T-E_S$ in denominator 
(in the latter case we perform a shift 
$\tilde\lambda_T=\lambda_T -i\pi T/3$) and proceed with the
analytical continuation $i\omega_n\to\omega+i 0^+$. 
Without loss of generality we assume $\omega > 0$.
As a result, we get for retarded (R) self-energies  
\begin{widetext}
\begin{equation}
Im \Sigma_{TST}^{(2)R}(\omega)\sim\left(J^{ST} \nu\right)^2
\int_{-D}^{D} d\xi_1\int_{-D}^{D} d\xi_2
n(\xi_2)(1-n(\xi_1))\delta(\omega +\xi_2 -\xi_1)
\label{eq8}
\end{equation}
\begin{equation}
Re  \Sigma_{TST}^{(2)R}(\omega)\sim\left(J^{ST} \nu\right)^2
 \int_{-D}^{D} d\xi_1  \int_{-D}^{D} d\xi_2
n(\xi_2)(1-n(\xi_1))P\frac{1}{\omega+\xi_2-\xi_1}
\label{eq9}
\end{equation}
\begin{equation}
Im \Sigma_{TTT}^{(2)R}(\omega)\sim\left(J^T \nu\right)^2
\int_{-D}^{D} d\xi_1\int_{-D}^{D} d\xi_2
n(\xi_2)(1-n(\xi_1))\delta(\omega +\xi_2 -\xi_1-\delta)
\label{eq8a}
\end{equation}
\begin{equation}
Re  \Sigma_{TTT}^{(2)R}(\omega)\sim\left(J^T \nu\right)^2
 \int_{-D}^{D} d\xi_1  \int_{-D}^{D} d\xi_2
n(\xi_2)(1-n(\xi_1))P\frac{1}{\omega+\xi_2-\xi_1-\delta}
\label{eq9a}
\end{equation}
\end{widetext}
where P denotes the principal value of the integral.

We start with discussion of self-energy parts 
determining the spin relaxation due to $T\to S$ transitions shown in  Fig. 5(a,b).
Assuming $T\ll D$ and neglecting temperature corrections at low temperatures 
$\omega \gg T$, we get
\begin{widetext}
\begin{equation}
Im \Sigma_{TST}^{(2)R}(\omega)\sim\left(J^{ST} \nu\right)^2\int_{0}^{D} d\xi_1
\int_{-D}^{0} d\xi_2
\delta(\omega +\xi_2 -\xi_1)
\sim\left(J^{ST} \nu(0)\right)^2\int_{0}^{\omega} d\xi 
\sim \left(J^{ST} \nu\right)^2\omega
\label{eq10}
\end{equation}
\begin{equation}
Re  \Sigma_{TST}^{(2)R}(\omega)\sim\left(J^{ST} \nu\right)^2
\int_{0}^{D} d\xi_1 \int_{-D}^{0} d\xi_2 P
\frac{1}{\omega+\xi_2-\xi_1}\sim\left(J^{ST} \nu\right)^2 \omega\ln\left(
\frac{D}{\omega}\right)
\label{eq11}
\end{equation}
\end{widetext}
In the opposite limit $T\gg \omega$
\begin{equation}
Im \Sigma_{TST}^{(2)R}(\omega)\sim\left(J^{ST} \nu\right)^2 T,
\label{eq10a}
\end{equation}
\begin{equation}
Re \Sigma_{TST}^{(2)R}(\omega)\sim\left(J^{ST} \nu\right)^2 \omega\ln\left(
\frac{D\gamma}{2\pi T}\right)
\label{eq10b}
\end{equation}
where $\ln \gamma=C=0.577...$ is the Euler constant.

Next we turn to calculation of the triplet level damping 
due to TT relaxation processes (Fig. 5a,b).
According to the Feynman codex, we can put $E_S=0$ at the first stage since 
the population of triplet
excited state is controlled by finite level splitting $\delta$. 
The contribution from diagram Fig. 5a 
is given by
\begin{equation}
Im \Sigma^{(2 LL)}_{TTT}=Im \Sigma^{(2 RR)}_{TTT}
\sim \left( J^{T}_0\nu\right)^2 (\omega -\delta)\;\;
\theta(\omega-\delta)
\label{eq22}
\end{equation}
\begin{equation}
Re \Sigma^{(2 LL)}_{TTT}=
Re \Sigma^{(2 RR)}_{TTT}\sim \left( J^{T}_0\nu\right)^2 (\omega -\delta)\;\;
\ln\left|\frac{D}{\omega-\delta}\right|
\label{eq22c}
\end{equation}
Similarly for Fig.5b, 
\begin{equation}
Im \Sigma^{(2 LR)}_{TTT}=
Im \Sigma^{(2 RL)}_{TTT}\sim \left( J^{T}_0\nu\right)^2 (\omega -\delta)\;\;
\theta(\omega-\delta)
\label{eq22a}
\end{equation}
and, with logarithmic accuracy
\begin{equation}
Re \Sigma^{(2 LR)}_{TTT}=
Re \Sigma^{(2 RL)}_{TTT}\sim \left( J^{T}_0\nu\right)^2 (\omega -\delta)\;\;
\ln\left|\frac{D}{\omega-\delta}\right|
\label{eq22b}
\end{equation}
The threshold character of relaxation determined by the Fermi golden rule
is the source of 
 asymmetry in broadening of triplet line (see the text). 
\begin{figure*}
\includegraphics[width=0.2\textwidth]{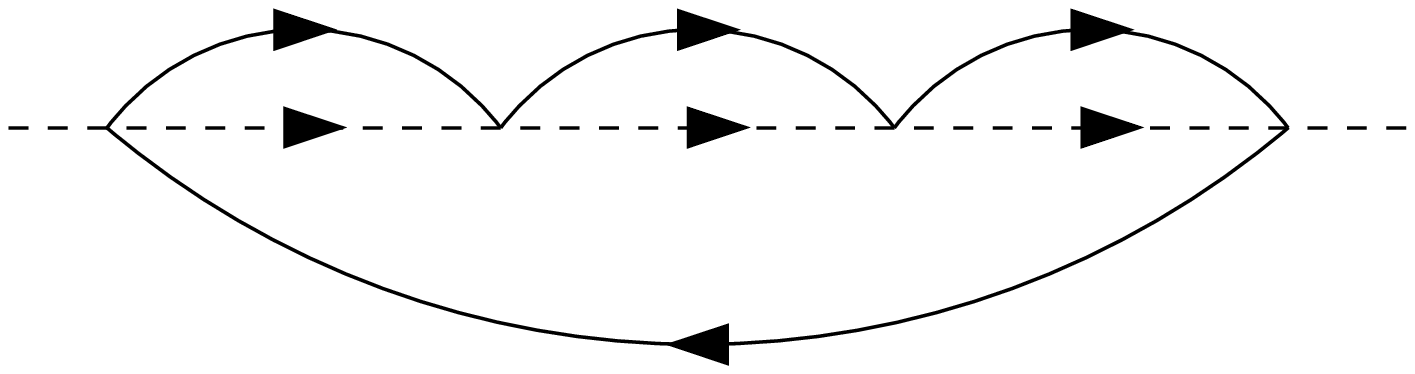}\hspace*{1cm}
\includegraphics[width=0.2\textwidth]{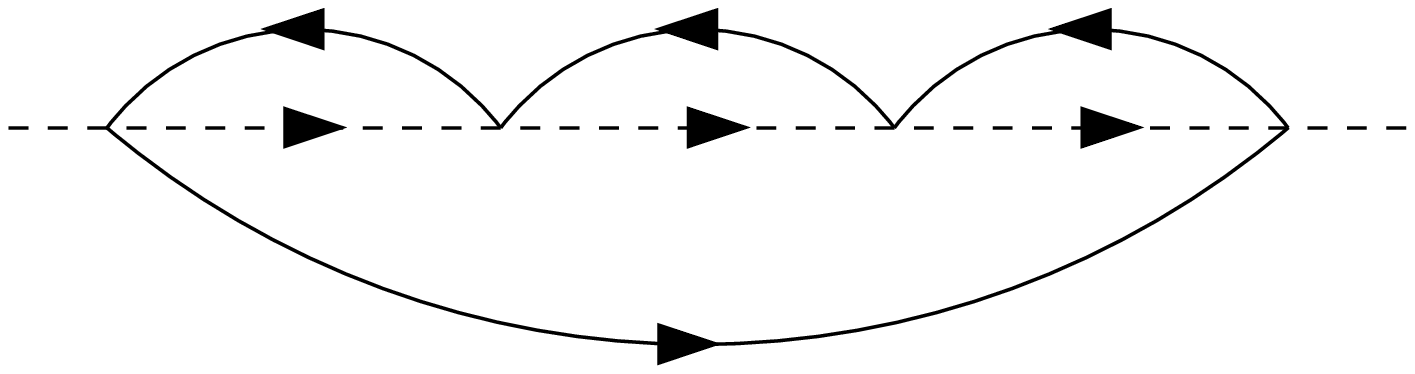}\hspace*{1cm}
\includegraphics[width=0.2\textwidth]{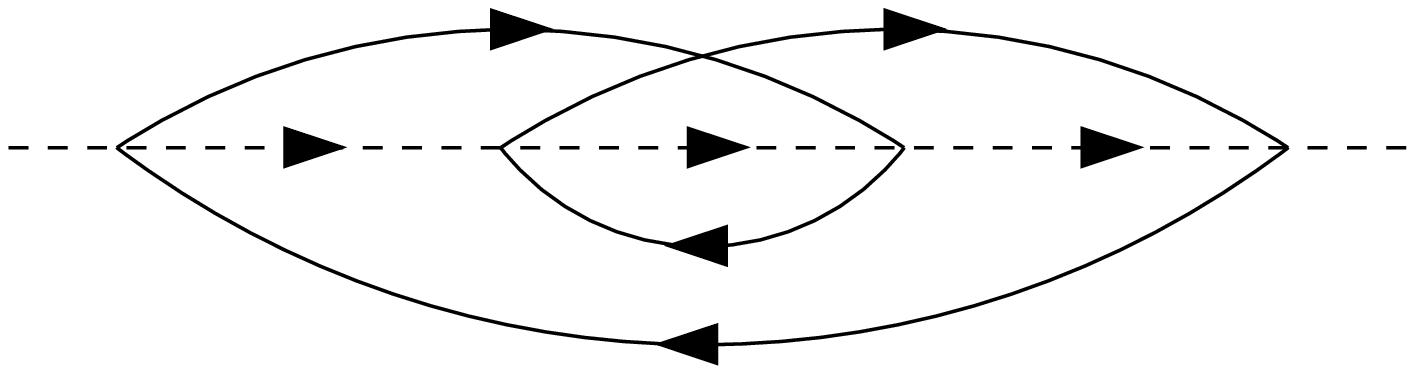}\\
\includegraphics[width=0.2\textwidth]{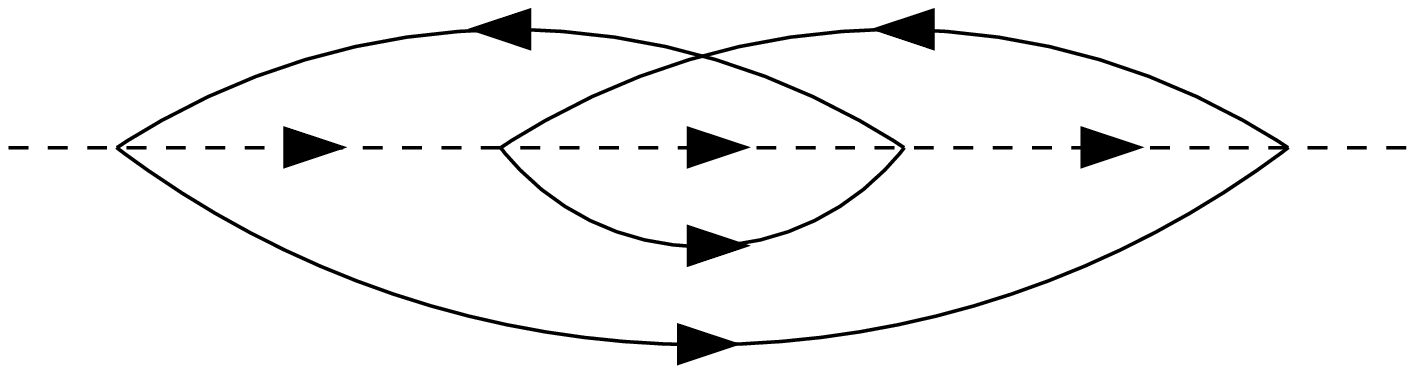}\hspace*{1cm}
\includegraphics[width=0.2\textwidth]{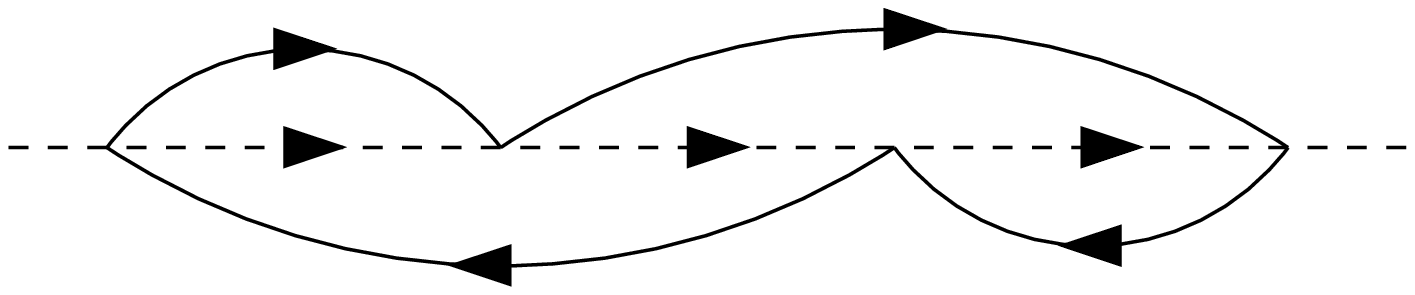}\hspace*{1cm}
\includegraphics[width=0.2\textwidth]{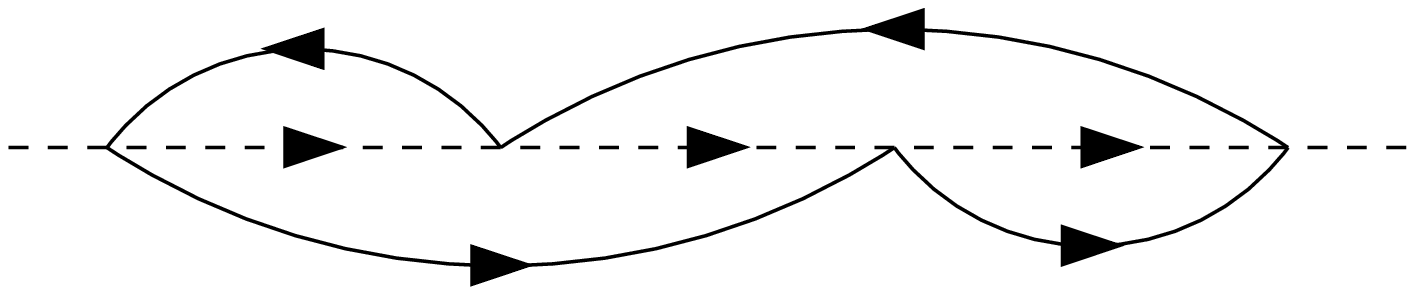} 
\caption{Fourth order leading diagrams (a-f)  for triplet self-energy part.}
\label{f2}
\end{figure*}
Now we turn to calculation of the third order diagrams 
$\Sigma^{(3)}$ shown in Fig. 5c,d.
\begin{widetext}
$$
\Sigma^{(3c)}(i\omega_n)\sim J^3 T^3
\sum_{\omega_{1,2,3}}\sum_{\bf k_{1,2,3}}G^0(-i\omega_1,-{\bf k}_1)
G^0(i\omega_2,{\bf k}_2)G^0(-i\omega_3,-{\bf k}_3)
{\cal G}^0(i\omega_n+i\omega_1+i\omega_2)
{\cal G}^0(i\omega_n+i\omega_2+i\omega_3)
$$
$$
\Sigma^{(3d)}(i\omega_n)\sim J^3 T^3
\sum_{\omega_{1,2,3}}\sum_{\bf k_{1,2,3}}G^0(i\omega_1,{\bf k}_1)
G^0(-i\omega_2,-{\bf k}_2)G^0(i\omega_3,{\bf k}_3)
{\cal G}^0(i\omega_n+i\omega_1+i\omega_2)
{\cal G}^0(i\omega_n+i\omega_2+i\omega_3)
$$
\end{widetext}
Evaluation of Matsubara sums gives
\begin{widetext}
\begin{equation}
\Sigma^{(3c)}(i\omega_n)\sim \left(J \nu\right)^3
\int_{-D}^{D} d\xi_1\int_{-D}^{D} d\xi_2\int_{-D}^{D} d\xi_3
\frac{n(\xi_2)(1-n(\xi_1))(1-n(\xi_3))}{(i\omega_n +\xi_2 -\xi_3-\lambda_1)
(i\omega_n +\xi_2 -\xi_1-\lambda_2)}
\label{eq12}
\end{equation}
\begin{equation}
\Sigma^{(3d)}(i\omega_n)\sim \left(J\nu\right)^3
\int_{-D}^{D} d\xi_1\int_{-D}^{D} d\xi_2\int_{-D}^{D} d\xi_3
\frac{n(\xi_1)n(\xi_3)(1-n(\xi_2))}{(i\omega_n +\xi_3 -\xi_2-\lambda_1)(i\omega_n +\xi_1 -\xi_2-\lambda_2)}
\label{eq13}
\end{equation}
\end{widetext}
Let us consider first the case $\lambda_1=\lambda_2=\lambda_S=0$ which corresponds to two 
singlet fermionic lines inserted in self-energy part.
Analytical continuation leads to following expression for $\Sigma^{(3)}=\Sigma^{(3b)}+\Sigma^{(3c)}$
at $T \ll \omega$
\begin{equation}
Im \Sigma_{TSST}^{(3)} (\omega)\sim\left(J^{ST}\nu\right)^3\frac{J^S}{J^{ST}}\left[\omega\ln\left(\frac{D}{\omega}\right)-\omega\right]
\label{eq14}
\end{equation}
$$
Re \Sigma^{(3)}_{TSST}(\omega)\sim\left(J^{ST}\nu\right)^3\frac{J^S}{J^{ST}}\;
\omega \; Re \left[ Li_2\left(-\frac{D}{\omega}\right)\right] \sim
$$
\begin{equation}
\sim
\left(\frac{J}{D}\right)^3\omega\ln^2\left(\frac{D}{\omega}\right)
\end{equation}
where $Li_2(x)$ is a di-logarithm function \cite{abram}.
As we already noticed, the first log correction to $Im \Sigma$ 
appears only in 3-rd order of the perturbation theory.
Thus,
\begin{equation}
Im \Sigma_{TSST}(\omega)\sim\left(J^{ST} \nu\right)^2 \omega
\left[1+ a \left(J^S \nu\right)\ln\left(\frac{D}{\omega}\right)+ ...\right]
\label{eq15}
\end{equation}
\begin{widetext}
\begin{equation}
Re \Sigma_{TSST}(\omega)\sim\left(J^{ST} \nu\right)^2
\left[ \omega\ln\left|\frac{D}{\omega}\right|
\left\{1+ b \left(J^S \nu\right)\ln\left|\frac{D}{\omega}\right|+ ...\right\}
+c(\omega-\delta)\ln\left|\frac{D}{\omega-\delta}\right|
\left\{1+ d \left(J^S \nu\right)\ln\left|\frac{D}{\omega-\delta}\right|+ ...
\right\}\right]
\label{eq15a}
\end{equation}
\end{widetext}
with coefficient $a,b,c,d \sim 1$. This reproduces results of Abrikosov-Migdal theory \cite{abr70}. 

We assume now that $\lambda_1=\lambda_2=\lambda_T=\delta$.  
It corresponds to the situation when both internal semi-fermionic GF correspond to
different components of the triplet. Following the same routine as for
calculation of $\Sigma^{(2)}$ we find
\begin{widetext}
\begin{equation}
Im \Sigma_{TTTT}^{(3)} (\omega)\sim\left(J^{T}\nu\right)^3
\left[(\omega-\delta)\ln\left|\frac{D}{\omega-\delta}\right|-(\omega-\delta)\right]\theta\left(\omega-\delta\right)
\label{eq16}
\end{equation}
\end{widetext}
Thus, the corrections to the relaxation rate associated 
with transitions between different 
components of the triplet have a threshold character 
determined by the energy conservation.

Finally, we consider a possibility when two internal semi-fermionic GF correspond to different states,
e.g. $\lambda_1=\lambda_S=0$, whereas $\lambda_2=\lambda_T=\delta$. Performing the calculations, 
one finds
\begin{widetext}
\begin{equation}
Im \Sigma_{TSTT}^{(3)} (\omega)\sim\left(J^{ST}\nu\right)^3\frac{J^T}{J^{ST}}
\left(
\left[
\delta\ln\left|\frac{D}{\delta}\right|-(\delta-\omega)
\ln\left|\frac{D}{\delta-\omega}\right|-\omega\right]+
\left[\delta\ln\left|\frac{D}{\delta}\right|-\omega
\ln\left|\frac{D}{\omega}\right|-(\delta-\omega)\right]\theta\left(\omega-\delta\right)
\right)
\label{eq17}
\end{equation}
\end{widetext}
Similar expression can be derived for $Im \Sigma_{TTST}^{(3)} (\omega)$.

Any insertion of the triplet
line in diagrams Fig.5.(a-d) results in additional suppression of
corresponding contribution for $\omega< eV$, which, in turn,
prevents the effective renormalization of the vertex $J^{S}$ in
contrast to the processes shown in Fig. 3. 
The leading corrections in the 4-th order of perturbation theory are shown on
Fig. 6a-f. We point out that all corrections to 
$Im \Sigma^{(n\geq 2)} \sim \omega\ln^{n-2}(D/\omega)$, 
 $Re \Sigma^{(n\geq 2)} \sim \omega\ln^{n-1}(D/\omega)$ 
and contain an additional power of the small parameter $\delta/D \ll 1$ 
as $\omega \to \delta$.


\begin{thebibliography}{99}
\bibitem{add} Possibility  of additional Kondo peaks at $eV\gg T_K$ in strong magnetic field
was noticed also in papers of Ref. [18], [21].
\end{thebibliography}
\end{document}